\documentclass[%
 reprint,
superscriptaddress,
 amsmath,amssymb,
 aps,
 nolongbibliography,
 prl,
 nobalancelastpage
]{revtex4-2}

\usepackage{graphicx}
\usepackage{dcolumn}
\usepackage{bm}
\usepackage{hyperref}
\hypersetup{
	colorlinks   = true, 
	urlcolor     = blue, 
	linkcolor    = blue, 
	citecolor    = blue 
}

\usepackage{esint}
\usepackage{amsfonts}
\usepackage{gensymb}
\usepackage{siunitx}
\usepackage{verbatim}
\usepackage{graphicx}
\usepackage{textcomp}
\usepackage{times}
\usepackage{subcaption}
\usepackage{ragged2e}
\DeclareCaptionJustification{justified}{\justifying}
\usepackage{xcolor,colortbl}
\definecolor{LightCyan}{rgb}{0.88,1,1}
\definecolor{LightMagenta}{rgb}{1,0.88,1}
\newcolumntype{a}{>{\columncolor{LightCyan}}c}
\newcolumntype{b}{>{\columncolor{LightMagenta}}c}

\begin{document}

\title{Theorem for spin-1 photonic pairs and non-Abelian holonomy in Maxwell Hamiltonians}

\author{Todd Van Mechelen}
\altaffiliation{equal contribution}
\affiliation{Elmore Family School of Electrical and Computer Engineering, Birck Nanotechnology Center, Purdue University, West Lafayette, Indiana 47907, USA}

\author{Wenbo Sun}
\altaffiliation{equal contribution}
\affiliation{Elmore Family School of Electrical and Computer Engineering, Birck Nanotechnology Center, Purdue University, West Lafayette, Indiana 47907, USA}

\author{Zubin Jacob}
\email{zjacob@purdue.edu}
\affiliation{Elmore Family School of Electrical and Computer Engineering, Birck Nanotechnology Center, Purdue University, West Lafayette, Indiana 47907, USA}

\begin{abstract}
Kramers degeneracy of spin-1/2 electrons, enforced by time reversal $\mathcal T_e^{2}=-1$, protects electronic spin doublets that can support non-Abelian geometric evolution. In contrast, photons carry $\mathcal T_{ph}^{2}=+1$ that does not enforce the degeneracy, and previous works on photonic Kramers analogs largely focused on pseudospin doublets engineered in photonic crystals and metamaterials without non-Abelian geometric evolution. Here, we prove a Kramers theorem for spin-1 photons through the symmetries of Maxwell Hamiltonians. The key ingredient of our theory is the parity-time anti-commutativity, which enforces the emergence of spin-1 photonic pairs at the same frequency and momentum. We further introduce magnetoelectric media that provide effective photon mass while preserving the parity-time anti-commutativity. We demonstrate that these massive photon pairs can acquire non-Abelian geometric phases within their symmetry-protected degeneracy under adiabatic evolutions of Maxwell Hamiltonians, in analogy to their electronic counterparts. Our results establish the spin-1 photonic counterparts of electronic Kramers pairs and open a new route towards non-Abelian geometric control of light.
\end{abstract}

\maketitle 

\emph{Introduction.-}Electronic Kramers degeneracy originates from time-reversal symmetry for spin-1/2 electrons, where $\mathcal T_e^2=-1$ forces the electronic energy levels of a time reversal invariant system to be at least two-fold degenerate~\cite{wigner2012group}. This protected degeneracy plays a central role in the spin structure of atoms, molecules, and solids~\cite{mead1992geometric,hasan2010colloquium}. In inversion symmetric systems, time reversal and inversion together enforce the degeneracy of electronic bands at every momentum, giving rise to Kramers pairs with the same energy and momentum~\cite{fu2007topological,hasan2010colloquium,qi2011topological}. These spin-1/2 electronic Kramers pairs provide a natural subspace for non-Abelian geometric phases~\cite{berry1984quantal,wilczek1984appearance,xiao2010berry,avron1988topological,mead1987molecular}. Unlike the Abelian Berry phase of a nondegenerate state, their geometric evolution under adiabatic modulation of the Hamiltonian around a closed loop can mix the two partners and depends on the order of the adiabatic cycles. These associated non-Abelian geometric phases are central to holonomic quantum gates and spin transport in spintronic devices~\cite{murakami20042,sinova2015spin,zanardi1999holonomic,golovach2010holonomic,bernevig2005holonomic}.

In contrast, photons are spin-1 bosons with time reversal $\mathcal T_{ph}^2=+1$, which does not enforce Kramers degeneracy by itself. Previous works in photonics have largely mimicked Kramers pairs with pseudospin engineering and effective spin-1/2 theories, where selected photonic modes form a pseudospin doublet whose degeneracy is built into photonic structures, e.g., photonic crystal symmetries~\cite{khanikaev2013photonic,ozawa2019topological,chen2014experimental,wu2015scheme,slobozhanyuk2017three,he2016photonic,khanikaev2017two,chen2023photonic,mu2026spatial}. These constructions largely leveraged pseudospins that are conserved by design, and the two partners acquire separate Abelian Berry phases under adiabatic evolutions~\cite{khanikaev2013photonic,yang2024non}. Meanwhile, recent developments reveal the importance of non-Abelian geometric phases of light in non-Abelian topological photonics and holonomic photonic gates~\cite{yang2024non,song2025artificial,pinske2022geometrically,neef2025non,pachos2000optical,qin2026photonic}. Current realizations focus on engineered degeneracies in coupled waveguides, synthetic dimensions, and photonic lattices~\cite{chen2019non,sun2022non,pinske2022symmetry,neef2023three,chen2025high,cheng2025non,zhang2022non,brosco2021non,yang2019synthesis,yan2023non,polimeno2021experimental,kim2026programmable,guo2021experimental,iadecola2016non,parto2023non,lai2025non}. Spin-1 photonic counterparts of electronic Kramers pairs, degenerate at the same frequency and momentum and carrying non-Abelian geometric phases within the symmetry-enforced degeneracy, remain much less explored.

In this Letter, we prove a Kramers theorem for spin-1 photonic pairs in Maxwell Hamiltonians. The key ingredient of our theory is the parity-time anti-commutativity, which pairs photon states with an orthogonal partner at the same frequency and momentum. We further derive a class of magnetoelectric media that exhibit this parity-time anti-commutativity and introduce an effective photon mass, which opens a band gap while preserving the spin-1 photonic pairs. Unlike electronic Kramers pairs with the same parity, our spin-1 photonic pairs are labeled by opposite parity labels and acquire longitudinal polarization components controlled by effective mass. We further demonstrate that adiabatic modulation of Maxwell Hamiltonians along closed loops mixes these massive photonic pairs through noncommuting geometric transformations, revealing non-Abelian evolution (Wilczek-Zee holonomy) within their symmetry-protected subspace. Our results establish the spin-1 photonic pairs as the bosonic counterparts of the electron pairs and open a route toward non-Abelian geometric control of light.

\begin{figure}
    \centering
    \includegraphics[width= 3 in]{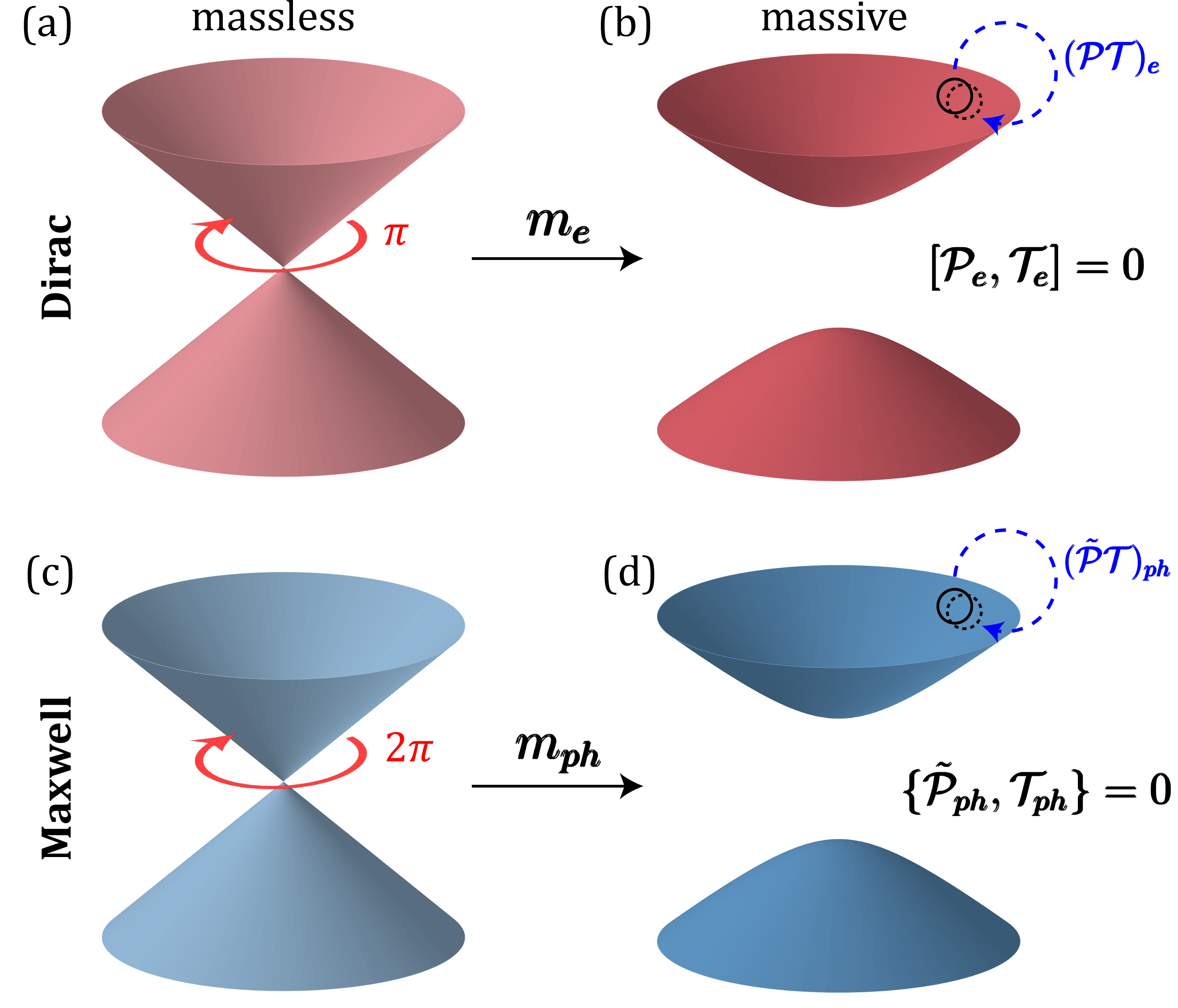}
    \caption{Dirac-Maxwell correspondence for electronic and photonic Kramers pairs. (a,b) Massless and massive spin-1/2 Dirac electrons. Time reversal $\mathcal{T}_e^2=-1$ and parity-time commutativity $[\mathcal P_e,\mathcal T_e]=0$ enforce Kramers pairs at the same energy and momentum. (c,d) Massless and massive spin-1 Maxwell photons. Time reversal $\mathcal{T}_{ph}^2=1$ and parity-time anti-commutativity $\{\mathcal{\tilde{P}}_{ph},\mathcal T_{ph}\}=0$ enforce Kramers pairs at the same frequency and momentum. The electron mass $m_e$ in (b) and the magnetoelectric photon mass $m_{ph}$ in (d) open the gaps without breaking the symmetries.}
    \label{fig:fig1}
\end{figure}

\emph{Maxwell Hamiltonian in Vacuum.-}For completeness, we first summarize some key aspects of the Maxwell Hamiltonian and Dirac-Maxwell correspondence in vacuum~\cite{van2018photonic,van2018quantum}. The correspondence between Dirac's and Maxwell's equations is illuminated in the Riemann-Silberstein (RS) basis~\cite{bialynicki2013role,barnett2014optical}, which we utilize to guide our search for spin-1 photonic pairs. In the RS basis, $\Psi_{\pm}=(\mathbf{E}\pm i\mathbf{H})/\sqrt{2}$ are complex vector fields often regarded as the wave function of the photon. For fields following $e^{i\mathbf{k}\cdot \mathbf{r}-i\omega t}$, we can combine Maxwell's equations in vacuum into a first-order wave equation similar in form to the massless Dirac equation,
\begin{equation}\label{eq:ms1-vacuum}
\omega \Psi =\mathcal{H}_{\mathrm{ph}}^{0}\Psi ,\qquad \mathcal{H}_{\mathrm{ph}}^{0}=c\,\sigma_{z}\otimes (\mathbf{k}\cdot \mathbf{S}),
\end{equation}
where $\Psi=(\Psi_+,\Psi_-)^T$, and the Pauli matrices $\sigma_i$ act on the two RS vectors. For comparison, massless Dirac electrons follow $E_e\Phi_e=H_e^0\Phi_e$ with $H_e^0=\hbar c\,\sigma_z\otimes(\mathbf{k}\cdot\boldsymbol{\sigma})$. As shown in Fig.~\ref{fig:fig1}(a, c), both particles are massless in vacuum with linear dispersion, $\omega=ck$ and $E_e=\hbar ck$, where $k=|\mathbf{k}|$. However, the two particles carry different spins that decide symmetries required for the emergence of Kramers pairs. Photons are spin-1 particles described by the SO(3) generators $(S_i)_{jk}=-i\epsilon_{ijk}$, while electrons are spin-1/2 particles described by the SU(2) generators. 

\emph{Parity-Time Anti-Commutativity.-}Kramers degeneracy is enforced by symmetries of the Hamiltonians. For spin-1/2 electrons, time reversal and parity guarantee orthogonal Kramers partners at the same energy and momentum~\cite{fu2007topological}. We now derive the symmetry of the spin-1 Maxwell Hamiltonian for the emergence of photonic pairs. To this end, we focus on Maxwell Hamiltonians with both time-reversal symmetry $\mathcal{T}^{-1}H_{ph}(-\mathbf{k})\mathcal{T}=H_{ph}(\mathbf{k})$ and parity symmetry $\mathcal{P}^{-1}H_{ph}(-\mathbf{k})\mathcal{P}=H_{ph}(\mathbf{k})$, where $H_{ph}(\mathbf{k})$ is the Maxwell Hamiltonian uniquely defined when photons are coupled to continuous media. We first note the key difference between electrons and photons lies in time reversal between fermions and bosons,
\begin{gather}\label{eq:ms1-time-reversal}
\begin{gathered}
\mathcal{T}_{e}=(\mathcal{I}_{2}\otimes \sigma_{y})\mathcal{K},\quad \mathcal{T}_{e}^{2}=-1,
\\
\mathcal{T}_{\mathrm{ph}}=\mathcal{K},\quad \mathcal{T}_{\mathrm{ph}}^{2}=+1,
\end{gathered}
\end{gather}
where $\mathcal{K}$ is the complex conjugation. The non-trivial phase $\mathcal{T}_{e}^2=e^{i\pi}=-1$ arises from the half-integer spin of the electron and is absent for integer spin photons. Therefore, $\mathcal{T}_{e}$ enforces electronic Kramers pairs, while $\mathcal{T}_{\mathrm{ph}}$ does not. Meanwhile, parity is trivial for both particles, $\mathcal{P}_{e}^{2}=\mathcal{P}_{\mathrm{ph}}^{2}=+1$. For electrons, parity and time reversal commute, $[\mathcal{P}_{e},\mathcal{T}_{e}]=0$, so that their combination maintains the non-trivial phase $(\mathcal{P}\mathcal{T})_{e}^{2}=-1$ and leads to the degenerate electronic Kramers pairs with the same frequency and momentum. In analogy, we argue that  photons can acquire this nontrivial phase if a parity operator $\tilde{\mathcal{P}}_{\mathrm{ph}}$ anti-commutes with time reversal,
\begin{equation}
\label{eq:ms1-anticommutation}
\{\tilde{\mathcal{P}}_{\mathrm{ph}},\mathcal{T}_{\mathrm{ph}}\}=0,\qquad (\tilde{\mathcal{P}}\mathcal{T})_{\mathrm{ph}}^{2}=-1,
\end{equation}
which is possible because $\mathcal{T}_{\mathrm{ph}}$ is anti-linear. We refer to Eq.~\eqref{eq:ms1-anticommutation} as the parity-time anti-commutativity. Hence, our search for spin-1 photonic pairs is directed at the parity operator $\tilde{\mathcal{P}}_{\mathrm{ph}}$ and the class of photonic media exhibiting this parity-time anti-commutativity.

Since $\mathcal{T}_{\mathrm{ph}}=\mathcal{K}$ is simply the complex conjugation in the RS basis, $\tilde{\mathcal{P}}_{\mathrm{ph}}$ must be purely imaginary. For concreteness, we now construct its explicit form in the RS basis. We identify a single operator that anti-commutes with $\mathcal{T}_{\mathrm{ph}}$ and leaves the vacuum Maxwell Hamiltonian invariant, $[(\tilde{\mathcal{P}}\mathcal{T})_{\mathrm{ph}},\mathcal{H}_{\mathrm{ph}}^{0}]=0$,
\begin{equation}
\label{eq:ms1-dressed-parity}
\tilde{\mathcal{P}}_{\mathrm{ph}}=\sigma_{y}\otimes \mathcal{I}_{3},\qquad \tilde{\mathcal{P}}_{\mathrm{ph}}^{*}=-\tilde{\mathcal{P}}_{\mathrm{ph}},\qquad \tilde{\mathcal{P}}_{\mathrm{ph}}^{2}=1,
\end{equation}
where $\mathcal{I}_{3}$ is the $3\times 3$ identity. Here, $\tilde{\mathcal{P}}_{\mathrm{ph}}=\mathcal{D}_{\mathrm{ph}}\mathcal{P}_{\mathrm{ph}}$ combines the physical spatial inversion $\mathcal{P}_{\mathrm{ph}}=-\sigma_{x}\otimes \mathcal{I}_{3}$ with the electromagnetic duality transformation $\mathcal{D}_{\mathrm{ph}}=i\sigma_{z}\otimes \mathcal{I}_{3}$ in the RS basis. We note that related mirror-based parity-time-duality symmetries protect photonic transport against backscattering~\cite{silveirinha2017p,geva2025polarization}. Since $\tilde{\mathcal{P}}_{\mathrm{ph}}$ and $\mathcal{T}_{\mathrm{ph}}$ each reverse $\mathbf{k}$, $(\tilde{\mathcal{P}}\mathcal{T})_{\mathrm{ph}}$ preserves $\mathbf{k}$ and transforms the wave function as $\Psi (\mathbf{k})\to (\sigma_{y}\otimes \mathcal{I}_{3})\Psi^{*}(\mathbf{k})$, which enforces degenerate spin-1 photonic pairs as discussed below. 

\emph{Spin-1 Photonic Pairs.-}The existence of the $\mathcal{PT}$ anti-commutativity allows us to prove a generalized Kramers theorem for spin-1 photonic pairs, in analogy to the electronic Kramers theorem. Maxwell Hamiltonians with both symmetries commute with their combination, $[(\tilde{\mathcal{P}}\mathcal{T})_{\mathrm{ph}},H]=0$. Therefore, for a state $|\psi \rangle$ with frequency $\omega$ satisfying $H|\psi \rangle =\omega |\psi \rangle$, $(\tilde{\mathcal{P}}\mathcal{T})_{\mathrm{ph}}|\psi \rangle$ is also an eigenstate with the same $\omega$ and $\mathbf{k}$. If the two were the same state, they could differ at most by a U(1) phase, $(\tilde{\mathcal{P}}\mathcal{T})_{\mathrm{ph}}|\psi \rangle =e^{i\phi}|\psi \rangle$. Applying $(\tilde{\mathcal{P}}\mathcal{T})_{\mathrm{ph}}$ again conjugates the phase, because $\mathcal{T}_{\mathrm{ph}}$ is anti-linear, and gives $|\psi \rangle =(\tilde{\mathcal{P}}\mathcal{T})_{\mathrm{ph}}^{2}|\psi \rangle =-|\psi \rangle$, which contradicts $|\psi \rangle \neq 0$. Therefore, $|\psi \rangle$ and $(\tilde{\mathcal{P}}\mathcal{T})_{\mathrm{ph}}|\psi \rangle$ are two different states, and every photonic band is two-fold degenerate at every $\mathbf{k}$.

These degenerate states can be expressed in a natural parity-labeled basis. We can construct a state $|\psi_{\mathbf{k}}^{+}\rangle$ satisfying,
\begin{equation} \label{eq:ms1-parity-plus}
\tilde{\mathcal{P}}_{\mathrm{ph}}|\psi_{-\mathbf{k}}^{+}\rangle =\zeta |\psi_{\mathbf{k}}^{+}\rangle ,\qquad \zeta =\pm 1.
\end{equation}
Applying time-reversal symmetry, we acquire its degenerate partner $|\psi_{\mathbf{k}}^{-}\rangle$,
\begin{equation} \label{eq:ms1-parity-minus}
|\psi_{\mathbf{k}}^{-}\rangle =\mathcal{T}_{\mathrm{ph}}|\psi_{-\mathbf{k}}^{+}\rangle ,\qquad \tilde{\mathcal{P}}_{\mathrm{ph}}|\psi_{-\mathbf{k}}^{-}\rangle =-\zeta |\psi_{\mathbf{k}}^{-}\rangle ,
\end{equation}
where the opposite sign follows directly from the  anti-commuting relation $\{\tilde{\mathcal{P}}_{\mathrm{ph}},\mathcal{T}_{\mathrm{ph}}\}=0$. As a result, Eqs.~\eqref{eq:ms1-parity-plus} and~\eqref{eq:ms1-parity-minus} show that $|\psi_{\mathbf{k}}^{\pm}\rangle =\mathcal{T}_{\mathrm{ph}}|\psi_{-\mathbf{k}}^{\mp}\rangle$ are spin-1 photonic partners with opposite parity labels $\pm \zeta$ through the inversion sewing relations, $\tilde{\mathcal{P}}_{\mathrm{ph}}|\psi_{-\mathbf{k}}^{\pm}\rangle =\pm \zeta |\psi_{\mathbf{k}}^{\pm}\rangle$. Furthermore, from the anti-unitarity of $\mathcal{T}_{\mathrm{ph}}$, we also have $\langle \psi_{\mathbf{k}}^{+}|\psi_{\mathbf{k}}^{-}\rangle =-\langle \psi_{-\mathbf{k}}^{+}|\psi_{-\mathbf{k}}^{-}\rangle =-\langle \mathcal{T}_{\mathrm{ph}}\psi_{\mathbf{k}}^{-}|\mathcal{T}_{\mathrm{ph}}\psi_{\mathbf{k}}^{+}\rangle =-\langle \psi_{\mathbf{k}}^{+}|\psi_{\mathbf{k}}^{-}\rangle =0$, which forces the two partners to be orthogonal, emphasizing the convenience of this
basis. 

Therefore, $|\psi_{\mathbf{k}}^{\pm}\rangle$ are the spin-1 photonic pairs. We note that $|\psi_{\mathbf{k}}^{\pm}\rangle$ are distinguished by opposite parity labels $\pm \zeta$. This is in stark contrast to electronic Kramers pairs, where $[\mathcal{P}_e,\mathcal{T}_e]=0$ forces the two partners to share the same parity labels through the inversion sewing relations, $\mathcal{P}_e|\phi_{-\mathbf{k}}^{\pm}\rangle = + \zeta |\phi_{\mathbf{k}}^{\pm}\rangle$ and are distinguished by opposite spin $\pm 1/2$ following the Dirac Hamiltonian $H_e$. At $\mathbf{k}=0$, the spin-1 photonic partners are $\mathcal{\tilde{P}}_{ph}$ eigenstates with eigenvalues $\pm 1$, whereas electronic Kramers partners share the same parity eigenvalue~\cite{fu2007topological}. Meanwhile, $(\tilde{\mathcal{P}}\mathcal{T})_{\mathrm{ph}}$ reverses the photon spin operator $\boldsymbol{\Sigma}=\mathcal{I}_{2}\otimes \mathbf{S}$, where $\mathbf S=(S_x,S_y,S_z)$ contains the spin-1 generators $S_i$. The pair thus carries opposite spin expectation values, $\mathbf{s}[\psi_{\mathbf{k}}^{-}]=-\mathbf{s}[\psi_{\mathbf{k}}^{+}]$ with $\mathbf{s}[\psi ]=\langle \psi |\boldsymbol{\Sigma}|\psi \rangle /\langle \psi |\psi \rangle$, although both partners remain spin-1 states and $\mathbf{s}$ is generally not quantized.

\emph{Parity-Time Anti-Commuting Media.-}We now explore the complete class of photonic media that preserves this parity-time anti-commutativity and gives rise to spin-1 photonic pairs. We consider the constitutive relations and the corresponding Maxwell's equations in the RS basis,
\begin{equation} \label{eq:ms1-constitutive}
\begin{pmatrix}
\mathbf{D} \\
\mathbf{B}
\end{pmatrix}=\begin{pmatrix}
\varepsilon & i\eta \\
-i\eta^{\dagger} & \mu 
\end{pmatrix}\begin{pmatrix}
\mathbf{E} \\
\mathbf{H}
\end{pmatrix},
\quad \ \
\omega \mathcal{M}(\omega ,\mathbf{k})\Psi =\mathcal{H}_{\mathrm{ph}}^{0}(\mathbf{k})\Psi ,
\end{equation}
where $\varepsilon =\varepsilon^{\dagger}$ and $\mu =\mu^{\dagger}$ are the $3\times 3$ permittivity and permeability tensors, and $\eta$ is an arbitrary $3\times 3$ magnetoelectric tensor that accounts for the cross-coupling of the electromagnetic fields. $\mathcal{M}$ is the response function of the material in the RS basis~\cite{silveirinha2015chern,van2018quantum} and can be represented in the block matrix form,
\begin{equation}
\mathcal{M}=\begin{bmatrix}
\alpha + \xi & \beta -i\gamma \\ \beta +i\gamma & \alpha - \xi
\end{bmatrix}=\mathcal{I}_2\otimes\alpha+\sigma_x\otimes\beta+\sigma_y\otimes\gamma+\sigma_z\otimes\xi,
\end{equation}
where $\mathcal{I}_2$ is the $2\times 2$ identity matrix, and the $3\times 3$ parameters are expressed in terms of the original material response, $\alpha =(\varepsilon +\mu )/2$, $\beta =(\varepsilon -\mu )/2$, $\gamma =(\eta -\eta^{\dagger})/(2i)$, and $\xi =(\eta +\eta^{\dagger})/2$. Here, $\xi$ and $i\gamma$ are the Hermitian and anti-Hermitian parts of magnetoelectric response, respectively. From Eq.~\eqref{eq:ms1-constitutive}, we can define the Maxwell Hamiltonian $\mathcal{H}_{\mathrm{ph}}(\mathbf{k})$ in arbitrary continuous media, 
\begin{equation}\label{eq:generalHME}
    \mathcal{H}_{\mathrm{ph}}(\mathbf{k})\Psi = [\mathcal{H}_{\mathrm{ph}}^{0}(\mathbf{k})-\omega\mathcal{M}(\omega ,\mathbf{k})+\omega\mathcal{I}_6]\Psi = \omega \Psi.
\end{equation}

Since $\mathcal{H}_{\mathrm{ph}}^{0}$ commutes with $(\tilde{\mathcal{P}}\mathcal{T})_{\mathrm{ph}}$, the medium supports spin-1 photonic pairs if the response function also commutes, $[(\tilde{\mathcal{P}}\mathcal{T})_{\mathrm{ph}},\mathcal{M}]=0$. We find the only allowable material responses are,
\begin{equation} \label{eq:ms1-compatible-media}
\alpha =\alpha^{T},\qquad \beta =\beta^{i}S_{i},\qquad \gamma =\gamma^{i}S_{i},\qquad \xi =\xi^{i}S_{i},
\end{equation}
where $\alpha$ and the coefficients $\beta^{i},\gamma^{i},\xi^{i}$ are real and $(S_j)_{ik}=-i\epsilon_{ijk}$, i.e., all response tensors except $\alpha$ must be antisymmetric. Here, Eq.~\eqref{eq:ms1-compatible-media} is equivalent to $\mu =\varepsilon^{*}$ and $\eta^{T}=-\eta$, and spatially local $\mathcal{T}_{\mathrm{ph}}$-invariant media thus fulfill Eq.~\eqref{eq:ms1-compatible-media} only if $\mu =\varepsilon$ and the magnetoelectric response $\eta =i\gamma$ is antisymmetric. We note that the condition $\mu =\varepsilon$ is too strict, and we generalize the parity-time anti-commutativity condition to $\varepsilon$ proportional to $\mu$ by using normalized RS vectors $\Psi_{\pm}=(\sqrt{\varepsilon }\mathbf{E}\pm i\sqrt{\mu }\mathbf{H})/\sqrt{2}$~\cite{bialynicki2013role} in the Supplemental Material. 

Meanwhile, we note that frequency and spatial dispersion (nonlocality) constraints naturally arise from symmetry constraints. Under $\mathcal{T}_{\mathrm{ph}}$ symmetry, response tensors satisfy $\alpha(\mathbf{k})=\alpha(-\mathbf{k})$ and $\gamma(\mathbf{k})=\gamma(-\mathbf{k})$ are even, while $\beta(\mathbf{k})=-\beta(-\mathbf{k})$ and $\xi(\mathbf{k})=-\xi(-\mathbf{k})$ are odd. Meanwhile, the reality of the electromagnetic fields further requires $\alpha(\omega)=\alpha(-\omega)$ and $\beta(\omega)=\beta(-\omega)$, while $\gamma(\omega)=-\gamma(-\omega)$ and $\xi(\omega)=-\xi(-\omega)$. The fact that $\gamma$ is the only element that is even in $\mathbf{k}$ while being odd in $\omega$ is important. As shown below, this contribution acts as an \textit{effective mass} for the photon, and will always open a bandgap in the medium regardless of polarization.

\begin{figure}
    \centering
    \includegraphics[width=3.4 in]{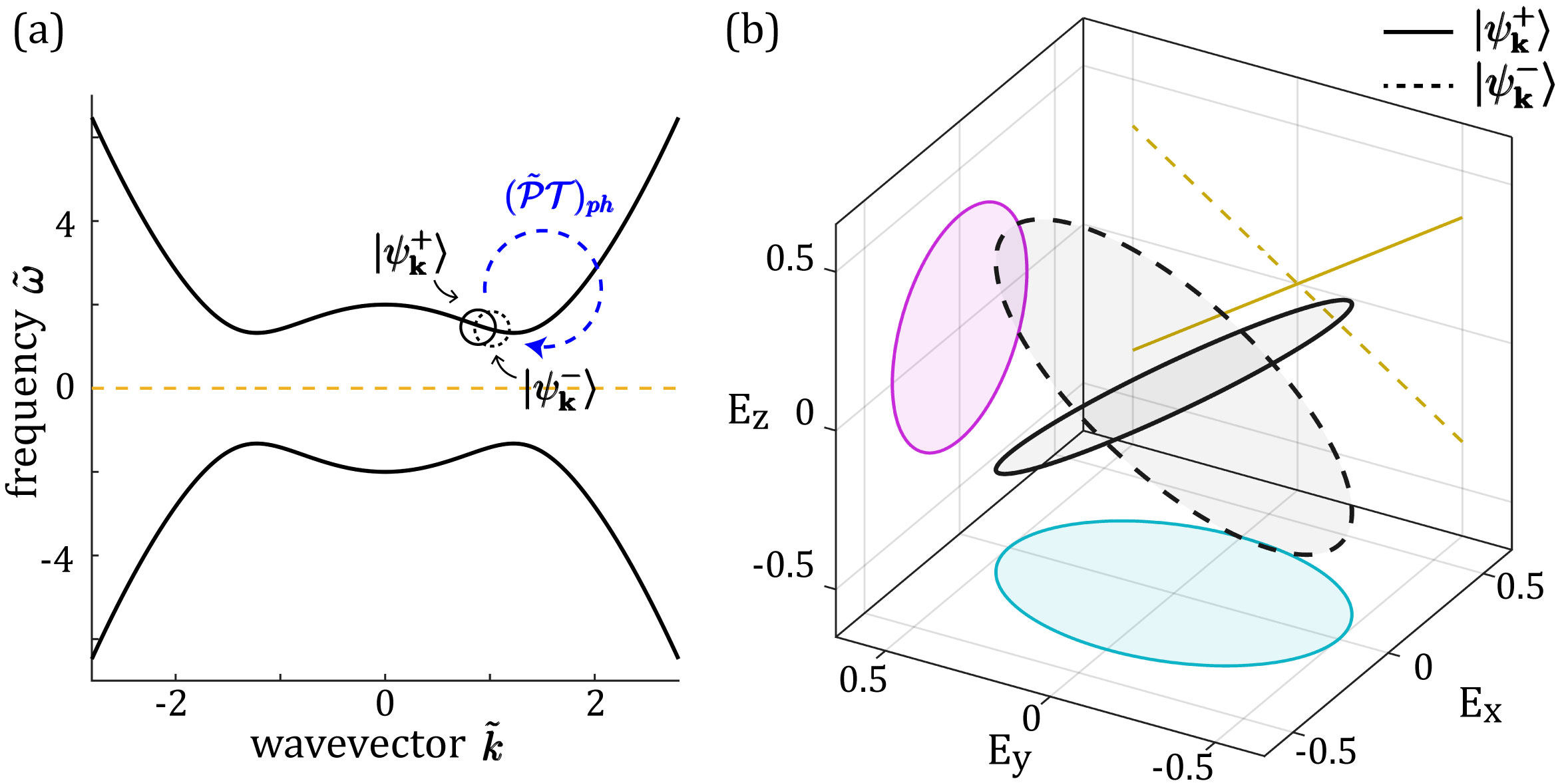}
    \caption{Massive spin-1 photonic pairs in the magnetoelectric medium. (a) Bulk dispersion $\omega=\pm v\sqrt{k^2+m_{ph}^2}$. Each branch is two-fold degenerate with spin-1 photonic pairs $\psi^\pm_{\mathbf k}$. The dashed line marks the zero-frequency pairs. $\tilde{k}=2k/m_0$ and $\tilde{\omega}=2\omega/vm_0$. (b) Electric field polarizations of the spin-1 photonic pair for $\psi^+_{\mathbf k}$ (solid) and $\psi^-_{\mathbf k}$ (dashed) at $\mathbf k=m_{ph} \hat{\mathbf x}$, with projections onto the three coordinate planes. The partners share the xy and xz projections. The effective photon mass $m_{ph}$ adds a longitudinal field component along $\mathbf k$.}
    \label{fig:fig2}
\end{figure}

\emph{Effective Magnetoelectric Photon Mass.-}For concreteness, we now demonstrate the spin-1 photonic pairs and introduce 2D magnetoelectric media with antisymmetric magnetoelectric response as the example media. We show that this antisymmetric magnetoelectric response provides an effective photon mass without breaking the spin-1 photonic pairs degeneracy. Following Eq.~\eqref{eq:ms1-compatible-media}, we consider a two-dimensional (2D) magnetoelectric medium with $\mathbf{k}=k_{x}\hat{\mathbf{x}}+k_{y}\hat{\mathbf{y}}$, isotropic positive $\varepsilon$ and $\mu$, and the antisymmetric magnetoelectric response $\gamma =\gamma^{z}(\omega ,\mathbf{k})S_{z}$, i.e., $\mathbf{D}=\varepsilon \mathbf{E}-\gamma^{z}S_{z}\mathbf{H}$ and $\mathbf{B}=\mu \mathbf{H}-\gamma^{z}S_{z}\mathbf{E}$. We find the Maxwell Hamiltonian,
\begin{equation}
\label{eq:ms1-mass-hamiltonian}
\mathcal{H}_{ph}^{ME}\Psi=\omega\Psi,\qquad \mathcal{H}_{ph}^{ME}=v\left[\frac{\mathcal{H}_{\mathrm{ph}}^0}{c}+m_{\mathrm{ph}}\,\sigma_y\otimes S_z\right],
\end{equation}
where $v=c/\sqrt{\varepsilon\mu}$ is the speed of light in the medium. Comparing Eq.~\eqref{eq:ms1-mass-hamiltonian} with the massive Dirac Hamiltonian $H_e=H_e^0+m_ec^2\,\sigma_x\otimes \mathcal{I}_2$, we immediately see that the photon acquires an effective mass,
\begin{equation}
\label{eq:ms1-photon-mass}
m_{\mathrm{ph}}(\omega,\mathbf{k})=-\frac{\omega}{c}\gamma^z(\omega,\mathbf{k}).
\end{equation}
This effective photon mass parameter $m_{\mathrm{ph}}$ yields a dispersion analogous to that of the electron mass $m_e$ in the Dirac equation,
\begin{equation}
\label{eq:ms1-dispersion}
\omega=v\sqrt{k^2+m_{\mathrm{ph}}^2},\qquad E_e=\sqrt{(\hbar ck)^2+(m_ec^2)^2}.
\end{equation}
Equation~\eqref{eq:ms1-dispersion} shows that $m_{\mathrm{ph}}$ opens a band gap at $\mathbf{k}=0$ [Fig.~\ref{fig:fig1}(b, d)]. In the following, for concreteness, we consider the minimal nonlocal model allowed by symmetry constraints, $m_{ph}(\omega,\mathbf k)=m_0-m_2k^{2}$, where $m_0$ is the local magnetoelectric response and the quadratic term represents the magnetoelectric analog of the Hall viscosity that renders the Hall conductivity quadratically momentum dependent~\cite{van2018quantum,van2021optical,sun2023optical}. Here, we have implicitly assumed  $m_{ph}$ to be frequency independent, which corresponds to $\gamma^z\propto 1/\omega$, consistent with $\gamma^z$ being odd in $\omega$, and can naturally arise as the high-frequency tail of an oscillator magnetoelectric resonance term, $m_{ph}(\omega)=m_0\,\omega^2/(\omega^2-\omega_0^2)$, at $\omega\gg\omega_0$. We note that our results depend only on the response at the operating frequency, and this simple $m_{ph}$ treatment preserves fundamental characteristics of response functions (e.g.,  positive definiteness and causality). In analogy to the spin-orbit interaction in graphene that opens a band gap without breaking time-reversal symmetry~\cite{kane2005quantum}, our viscous magnetoelectric $m_{\mathrm{ph}}$ mass term preserves $\tilde{\mathcal{P}}_{\mathrm{ph}}$ and $\mathcal{T}_{\mathrm{ph}}$. As a result, each band remains two-fold degenerate at every $\mathbf{k}$, as enforced by the parity-time anti-commutativity. 

\emph{Massive Spin-1 Photonic Pairs.-}We now demonstrate the wave functions of spin-1 photonic pairs in the viscous magnetoelectric medium. We define the bulk electromagnetic waves through the parity-labeled states $\psi^\pm_\mathbf{k}$ following Eqs.~(\ref{eq:ms1-parity-plus},~\ref{eq:ms1-parity-minus}). For the positive-frequency band in Eq.~\eqref{eq:ms1-dispersion} [Fig.~\ref{fig:fig2}(a)], we find,
\begin{equation}\label{eq:ms1-massive-pair}
\psi_{\mathbf{k}}^{\pm}=\frac{1}{2}\begin{pmatrix}-\frac{1}{q} (k\hat{\mathbf{z}}\pm m_{\mathrm{ph}}\hat{\mathbf{k}})+i\hat{\mathbf{t}}\\
\mp\frac{i}{q}(k\hat{\mathbf{z}}\mp m_{\mathrm{ph}}\hat{\mathbf{k}})\pm\hat{\mathbf{t}}\end{pmatrix},
\end{equation}
where $q=\sqrt{k^{2}+m_{\mathrm{ph}}^{2}}=\omega /v$, $\hat{\mathbf{k}}=\mathbf{k}/k$, and $\hat{\mathbf{t}}=\hat{\mathbf{z}}\times \hat{\mathbf{k}}$ is the in-plane unit vector transverse to $\mathbf{k}$. From Eq.~\eqref{eq:ms1-massive-pair}, $\psi_{\mathbf{k}}^{\pm}=\mathcal{T}_{\mathrm{ph}}\psi_{-\mathbf{k}}^{\mp}$ confirms that $\psi^\pm_\mathbf{k}$ are orthonormal time-reversal partners with opposite parity labels $\tilde{\mathcal{P}}_{\mathrm{ph}}\psi_{-\mathbf{k}}^{\pm}=\pm \psi_{\mathbf{k}}^{\pm}$ [Fig.~\ref{fig:fig2}(a)], and reduce to transverse plane waves in vacuum when $m_{\mathrm{ph}}=0$. 

In Fig.~\ref{fig:fig2}(b), we present the unique polarization of massive spin-1 photonic pairs. Here, $m_{\mathrm{ph}}$ modifies the transversality condition to $i\mathbf{k}\cdot \Psi_{\pm}=m_{\mathrm{ph}}\hat{\mathbf{z}}\cdot \Psi_{\mp}$, which adds a longitudinal component to the propagating fields, $\hat{\mathbf{k}}\cdot \mathbf{E}=-i(m_{\mathrm{ph}}/k)E_{z}$ and $\hat{\mathbf{k}}\cdot \mathbf{H}=i(m_{\mathrm{ph}}/k)H_{z}$, absent for plane waves in vacuum and conventional isotropic media. In Fig.~\ref{fig:fig2}(b), we plot the polarization ellipses of the two partners for $\mathbf k = \frac{m_0}{2} \hat{\mathbf x}$, where we take $m_0>0$, $m_2=2/m_0$, and the x component represents the longitudinal component of the propagating fields. We note that, as $k\to\infty$, the partners approach opposite circular polarizations. Meanwhile, the two partners carry opposite spin $\mathbf{s}[\psi_{\mathbf{k}}^{\pm}]=\mp (m_{\mathrm{ph}}/q)\hat{\mathbf{z}}$. 

In addition, we note that $\mathcal{H}_{ph}^{\mathrm{ME}}$ also possesses two degenerate zero-frequency states,
\begin{equation}
\label{eq:ms1-zero-pair}
\psi_{\mathbf{k}}^{0\pm}=\frac{1}{\sqrt{2}\, q}
\begin{pmatrix}
\mp \mathbf{k}+m_{\mathrm{ph}}\hat{\mathbf{z}} \\
i(\mathbf{k}\pm m_{\mathrm{ph}}\hat{\mathbf{z}})
\end{pmatrix},
\end{equation}
which form flat bands $\omega =0$ absent for Dirac electrons. We note that they are also time-reversal partners with opposite parity $\tilde{\mathcal{P}}_{\mathrm{ph}}\psi_{-\mathbf{k}}^{0\pm}=\pm \psi_{\mathbf{k}}^{0\pm}$ and zero spin $\mathbf{s}[\psi^{0\pm}_{\mathbf{k}}]=0$.

\begin{figure}[!t]
    \centering
    \includegraphics[width=3 in]{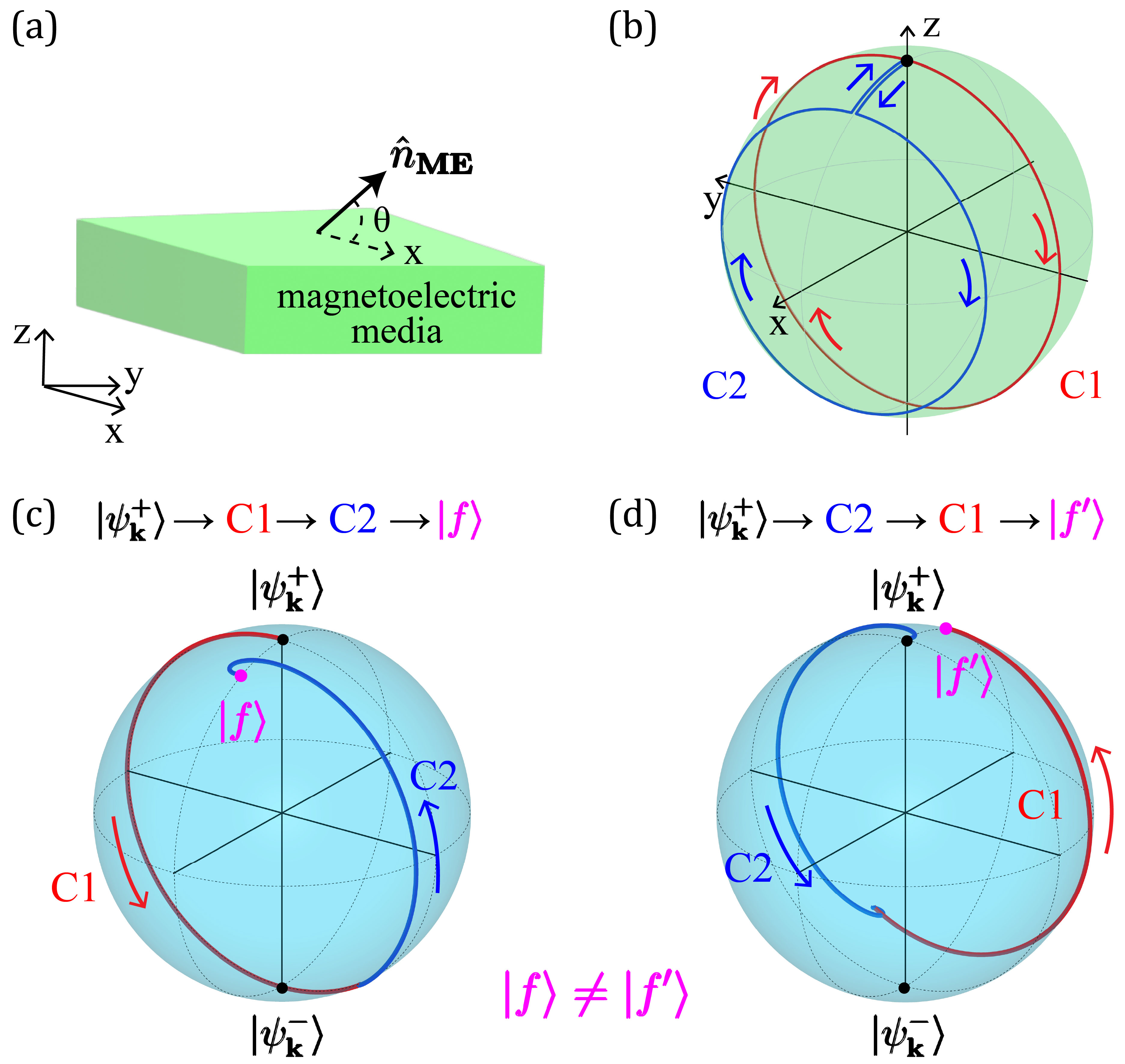}
    \caption{Non-Abelian Wilczek-Zee holonomy of massive spin-1 photonic pairs. (a) Two-dimensional magnetoelectric medium with in-plane wave vector $\mathbf k$ and magnetoelectric response axis $\hat{\mathbf n}$ at angle $\theta$ from $\mathbf k$. (b) Two cyclic adiabatic evolutions of the Maxwell Hamiltonian, where $\hat{\mathbf n}$ is rotated along closed paths from and back to $\hat{\mathbf z}$. (c,~d) Trajectories of the initial state $\psi^+_{\mathbf k}$ when the Maxwell Hamiltonian is evolved adiabatically along (c) $C_1$ and then $C_2$, or (d) along $C_2$ and then $C_1$. The Bloch sphere is spanned by the initial state pair $\psi^+_{\mathbf k}$ (north pole) and $\psi^-_{\mathbf k}$ (south pole), and each point on the Bloch sphere surface is the normalized projection onto the initial pair, with the common dynamical phase removed. The two evolution sequences end in different final states $|f\rangle \neq |f'\rangle$, revealing the non-Abelian nature of the geometric phase.}
    \label{fig:fig3}
\end{figure}

\emph{Non-Abelian Wilczek-Zee Holonomy.-}One unique feature of spin-1/2 electronic Kramers pairs at the same energy and momentum is the non-Abelian holonomy acquired by cyclic evolutions of Hamiltonians~\cite{avron1988topological,mead1987molecular}. Here, we reveal that the massive spin-1 photonic pairs can similarly acquire a symmetry-protected non-Abelian Wilczek-Zee (WZ) phase under cyclic modulation of Maxwell Hamiltonians. Unlike previous non-Abelian holonomies of light largely focusing on engineered degeneracies in waveguide systems~\cite{neef2023three,zhang2022non,sun2022non}, in our case, the degeneracy is enforced by the global symmetries in our time reversal invariant magnetoelectric medium.

To illustrate this idea concretely, in the following, we consider the adiabatic cyclic modulation of the Maxwell Hamiltonian through a rotation of the magnetoelectric response axis. Here, we replace $S_z$ in Eq.~\eqref{eq:ms1-mass-hamiltonian} by $\hat{\mathbf n}\cdot\mathbf S$, i.e., $\mathbf D=\varepsilon\mathbf E-\gamma^{z}(\hat{\mathbf n}\cdot\mathbf S)\mathbf H$ and $\mathbf B=\mu\mathbf H-\gamma^{z}(\hat{\mathbf n}\cdot\mathbf S)\mathbf E$, and consider rotating the axis $\hat{\mathbf n}$ adiabatically. From Eq.~\eqref{eq:ms1-compatible-media}, $\mathcal{H}_{\mathrm{ME}}$ commutes with $(\tilde{\mathcal{P}}\mathcal{T})_{\mathrm{ph}}$ for every $\hat{\mathbf{n}}$, and $\omega =vq$ does not depend on $\hat{\mathbf{n}}$. Therefore, the two partners remain degenerate along any path of $\hat{\mathbf{n}}$ and thus acquire no relative dynamical phase. 

We now consider the instantaneous spin-1 photonic pair $V(t)=(\psi_{\mathbf{k}}^{+}(t),\psi_{\mathbf{k}}^{-}(t))$ for the axis $\hat{\mathbf{n}}(t)$, which reduces to Eq.~\eqref{eq:ms1-massive-pair} at $\hat{\mathbf{n}}=\hat{\mathbf{z}}$. After removing the common dynamical phase, we find a closed cycle $C$ transforms the amplitudes of the two partners through~\cite{wilczek1984appearance,chen2025high},
\begin{equation} \label{eq:ms1-wz-holonomy}
W_{C}=\mathbb{P}_{\mathrm{path}}\exp\left( i\oint_{C} \mathcal{A}_{\mu}\, d\lambda^{\mu} \right),\quad \ \ \mathcal{A}_{\mu}=iV^{\dagger}\partial_{\mu}V,
\end{equation}
where $\mathbb{P}_{\mathrm{path}}$ denotes path ordering, $\mathcal{A}_{\mu}$ is the WZ connection, and $\lambda^{\mu}=(\theta ,\varphi )$ parametrize the axis $\hat{\mathbf{n}}=\cos\theta \, \hat{\mathbf{k}}+\sin\theta (\cos\varphi \, \hat{\mathbf{z}}-\sin\varphi \, \hat{\mathbf{t}})$. We find,
\begin{equation}
\mathcal{A}_{\theta}=(ab/2)\sin\theta \, \tau_{y}, \qquad \mathcal{A}_{\varphi}=a\tau_{x}-b\cos\theta \, \tau_{z},
\end{equation}
where $a=k/q$, $b=m_{\mathrm{ph}}/q$, and the Pauli matrices $\tau_{i}$ act on the two spin-1 photonic partner amplitudes. 

To explicitly demonstrate the non-Abelian nature of the geometric phase in the spin-1 photonic pairs, in the following, we consider two closed cycles acting on the same initial pair at $(\theta ,\varphi )=(\pi /2,0)$, i.e., $\hat{\mathbf{n}}=\hat{\mathbf{z}}$, as shown in Fig.~\ref{fig:fig3}(a,b). We first consider a cycle $C_1$ with $\theta =\pi /2$ and $\varphi = 0\to 2\pi$. We find the geometric phase acquired by the spin-1 photonic pairs after this cyclic operation,
\begin{equation}
\label{eq:ms1-equatorial-loop}
W_1=e^{i\Phi_{1}\tau_{x}}=\begin{pmatrix}
\cos\Phi_{1} & i\sin\Phi_{1} \\
i\sin\Phi_{1} & \cos\Phi_{1}
\end{pmatrix},\qquad \Phi_{1}=2\pi a.
\end{equation}
Equation~\eqref{eq:ms1-equatorial-loop} shows that one rotation of the magnetoelectric axis around $\mathbf{k}$ mixes the two partners of opposite parity and spin, and converts $|\psi_{\mathbf{k}}^{+}\rangle$ into $\cos\Phi_{1}|\psi_{\mathbf{k}}^{+}\rangle +i\sin\Phi_{1}|\psi_{\mathbf{k}}^{-}\rangle$. 

Meanwhile, we consider a second cycle $C_2$ as tilting the axis toward $\mathbf{k}$, rotating it, and then returning, i.e., $(\theta ,\varphi )=(\pi /2,0)\to (\theta_{1},0)\to (\theta_{1},2\pi )\to (\pi /2,2\pi )$ with $0<\theta_{1}<\pi /2$. We find the geometric phase,
\begin{equation}
\label{eq:ms1-tilted-loop}
W_{2}=e^{i\delta \tau_{y}}\, e^{i2\pi (a\tau_{x}-b \cos\theta_{1} \tau_{z})}\, e^{-i\delta \tau_{y}},
\end{equation}
where $\delta =ab \cos\theta_{1} /2$. Equation~\eqref{eq:ms1-tilted-loop} represents a rotation $W_{2}=\cos\Phi_{2}+i\sin\Phi_{2}(u_{x}\tau_{x}+u_{z}\tau_{z})$, where $\Phi_{2}=2\pi \sqrt{a^{2}+b^{2}\cos^2\theta_{1}}$, $u_x=(2\pi/\Phi_2)[a\cos(2\delta)+b\cos\theta_{1}\sin(2\delta)]$, and $u_z=(2\pi/\Phi_2)[a\sin(2\delta)-b\cos\theta_{1}\cos(2\delta)]$. We find that the two cycles do not commute, 
\begin{equation}
[W_1,W_2]=2i\sin\Phi_{1}\sin\Phi_{2}\, u_{z}\tau_{y}\neq 0,
\end{equation}
the final states will thus depend on the order of the two adiabatic cycles, proving the non-Abelian WZ holonomy within the Kramers degeneracy protected spin-1 photonic pairs.

In Fig.~\ref{fig:fig3}(c,~d), we plot the trajectories of the evolution of the initial state $|\psi^+_{\mathbf k} \rangle$ when the Maxwell Hamiltonian is evolved adiabatically along C1 and then C2, or along C2 and then C1. Here, each point on the trajectory represents the normalized projection of the instantaneous spin-1 photonic pair $V(t)$ onto the initial pair $|\psi^{\pm}_{\mathbf k} \rangle$~\cite{sugawa2021wilson}, with the common dynamical phase removed. In Fig.~\ref{fig:fig3}(c,~d), we take $\mathbf k = \frac{m_0}{2} \hat{\mathbf x}$, $m_2=2/m_0$, and $\theta_{1}=\pi /3$. We find the two final states $|f\rangle$  and $|f'\rangle$ following the two evolution sequences are different, with $|\langle f|f'\rangle |^{2}\approx 0.85 <1$, which illustrates the non-Abelian nature of the geometric phases of the massive spin-1 photonic pairs. We emphasize that the parity-time anti-commutativity protects the degeneracy along the whole cycle without relying on accidental degeneracy. 

\emph{Conclusion.-}In summary, we have introduced spin-1 photonic pairs enforced by the parity-time anti-commutativity of Maxwell Hamiltonians. We revealed the global symmetry protected non-Abelian Wilczek-Zee holonomy in magneto-electric media. Our theorem can be potentially extended to other bosonic systems~\cite{yang2024non,peng2022phonons,fruchart2020dualities,yves2024symmetry} and open a new route towards non-Abelian geometric control of light.

\emph{Acknowledgments.-}This research was supported by the Office of Naval Research under award number N00014231270.


\bibliography{reference.bib}

@book{wigner2012group,
  title={Group theory: and its application to the quantum mechanics of atomic spectra},
  author={Wigner, Eugene},
  volume={5},
  year={2012},
  publisher={Elsevier}
}

@article{fu2007topological,
  title={Topological insulators with inversion symmetry},
  author={Fu, Liang and Kane, Charles L},
  journal={Physical Review B—Condensed Matter and Materials Physics},
  volume={76},
  number={4},
  pages={045302},
  year={2007},
  publisher={APS}
}

@article{mead1992geometric,
  title={The geometric phase in molecular systems},
  author={Mead, C Alden},
  journal={Reviews of modern physics},
  volume={64},
  number={1},
  pages={51},
  year={1992},
  publisher={APS}
}

@article{hasan2010colloquium,
  title={Colloquium: topological insulators},
  author={Hasan, M Zahid and Kane, Charles L},
  journal={Reviews of modern physics},
  volume={82},
  number={4},
  pages={3045--3067},
  year={2010},
  publisher={APS}
}

@article{qi2011topological,
  title={Topological insulators and superconductors},
  author={Qi, Xiao-Liang and Zhang, Shou-Cheng},
  journal={Reviews of modern physics},
  volume={83},
  number={4},
  pages={1057--1110},
  year={2011},
  publisher={APS}
}

@article{berry1984quantal,
  title={Quantal phase factors accompanying adiabatic changes},
  author={Berry, Michael Victor},
  journal={Proceedings of the Royal Society of London. A. Mathematical and Physical Sciences},
  volume={392},
  number={1802},
  pages={45--57},
  year={1984},
  publisher={The Royal Society, London}
}

@article{wilczek1984appearance,
  title={Appearance of gauge structure in simple dynamical systems},
  author={Wilczek, Frank and Zee, Anthony},
  journal={Physical Review Letters},
  volume={52},
  number={24},
  pages={2111},
  year={1984},
  publisher={APS}
}

@article{xiao2010berry,
  title={Berry phase effects on electronic properties},
  author={Xiao, Di and Chang, Ming-Che and Niu, Qian},
  journal={Reviews of modern physics},
  volume={82},
  number={3},
  pages={1959--2007},
  year={2010},
  publisher={APS}
}

@article{avron1988topological,
  title={Topological invariants in Fermi systems with time-reversal invariance},
  author={Avron, JE and Sadun, L and Segert, J and Simon, B},
  journal={Physical review letters},
  volume={61},
  number={12},
  pages={1329},
  year={1988},
  publisher={APS}
}

@article{mead1987molecular,
  title={Molecular Kramers degeneracy and non-Abelian adiabatic phase factors},
  author={Mead, C Alden},
  journal={Physical review letters},
  volume={59},
  number={2},
  pages={161},
  year={1987},
  publisher={APS}
}

@article{golovach2010holonomic,
  title={Holonomic quantum computation with electron spins in quantum dots},
  author={Golovach, Vitaly N and Borhani, Massoud and Loss, Daniel},
  journal={Physical Review A—Atomic, Molecular, and Optical Physics},
  volume={81},
  number={2},
  pages={022315},
  year={2010},
  publisher={APS}
}

@article{bernevig2005holonomic,
  title={Holonomic quantum computing based on the Stark effect},
  author={Bernevig, BA and Zhang, Shou-Cheng},
  journal={Physical Review B—Condensed Matter and Materials Physics},
  volume={71},
  number={3},
  pages={035303},
  year={2005},
  publisher={APS}
}

@article{zanardi1999holonomic,
  title={Holonomic quantum computation},
  author={Zanardi, Paolo and Rasetti, Mario},
  journal={Physics Letters A},
  volume={264},
  number={2-3},
  pages={94--99},
  year={1999},
  publisher={Elsevier}
}

@article{murakami20042,
  title={SU (2) non-Abelian holonomy and dissipationless spin current in semiconductors},
  author={Murakami, Shuichi and Nagosa, Naoto and Zhang, Shou-Cheng},
  journal={Physical Review B—Condensed Matter and Materials Physics},
  volume={69},
  number={23},
  pages={235206},
  year={2004},
  publisher={APS}
}

@article{sinova2015spin,
  title={Spin hall effects},
  author={Sinova, Jairo and Valenzuela, Sergio O and Wunderlich, J{\"o}rg and Back, Craig H and Jungwirth, Tom{\'a}{\v{s}}},
  journal={Reviews of modern physics},
  volume={87},
  number={4},
  pages={1213--1260},
  year={2015},
  publisher={APS}
}

@article{khanikaev2013photonic,
  title={Photonic topological insulators},
  author={Khanikaev, Alexander B and Hossein Mousavi, S and Tse, Wang-Kong and Kargarian, Mehdi and MacDonald, Allan H and Shvets, Gennady},
  journal={Nature materials},
  volume={12},
  number={3},
  pages={233--239},
  year={2013},
  publisher={Nature Publishing Group UK, London}
}

@article{ozawa2019topological,
  title={Topological photonics},
  author={Ozawa, Tomoki and Price, Hannah M and Amo, Alberto and Goldman, Nathan and Hafezi, Mohammad and Lu, Ling and Rechtsman, Mikael C and Schuster, David and Simon, Jonathan and Zilberberg, Oded and others},
  journal={Reviews of Modern Physics},
  volume={91},
  number={1},
  pages={015006},
  year={2019},
  publisher={APS}
}

@article{chen2014experimental,
  title={Experimental realization of photonic topological insulator in a uniaxial metacrystal waveguide},
  author={Chen, Wen-Jie and Jiang, Shao-Ji and Chen, Xiao-Dong and Zhu, Baocheng and Zhou, Lei and Dong, Jian-Wen and Chan, Che Ting},
  journal={Nature communications},
  volume={5},
  number={1},
  pages={5782},
  year={2014},
  publisher={Nature Publishing Group UK, London}
}

@article{wu2015scheme,
  title={Scheme for achieving a topological photonic crystal by using dielectric material},
  author={Wu, Long-Hua and Hu, Xiao},
  journal={Physical review letters},
  volume={114},
  number={22},
  pages={223901},
  year={2015},
  publisher={APS}
}

@article{slobozhanyuk2017three,
  title={Three-dimensional all-dielectric photonic topological insulator},
  author={Slobozhanyuk, Alexey and Mousavi, S Hossein and Ni, Xiang and Smirnova, Daria and Kivshar, Yuri S and Khanikaev, Alexander B},
  journal={Nature Photonics},
  volume={11},
  number={2},
  pages={130--136},
  year={2017},
  publisher={Nature Publishing Group UK, London}
}

@article{he2016photonic,
  title={Photonic topological insulator with broken time-reversal symmetry},
  author={He, Cheng and Sun, Xiao-Chen and Liu, Xiao-Ping and Lu, Ming-Hui and Chen, Yulin and Feng, Liang and Chen, Yan-Feng},
  journal={Proceedings of the National Academy of Sciences},
  volume={113},
  number={18},
  pages={4924--4928},
  year={2016},
  publisher={National Academy of Sciences}
}

@article{khanikaev2017two,
  title={Two-dimensional topological photonics},
  author={Khanikaev, Alexander B and Shvets, Gennady},
  journal={Nature photonics},
  volume={11},
  number={12},
  pages={763--773},
  year={2017},
  publisher={Nature Publishing Group UK, London}
}

@article{chen2023photonic,
  title={Photonic Dirac cavities with spatially varying mass term},
  author={Chen, Kai and Komissarenko, Filipp and Smirnova, Daria and Vakulenko, Anton and Kiriushechkina, Svetlana and Volkovskaya, Irina and Guddala, Sriram and Menon, Vinod and Al{\`u}, Andrea and Khanikaev, Alexander B},
  journal={Science advances},
  volume={9},
  number={12},
  pages={eabq4243},
  year={2023},
  publisher={American Association for the Advancement of Science}
}

@article{mu2026spatial,
  title={Spatial Inversion Kramers Degeneracy},
  author={Mu, Jialu and Yang, Biao and Guo, Qinghua},
  journal={Physical Review Letters},
  volume={136},
  number={13},
  pages={136601},
  year={2026},
  publisher={APS}
}

@article{yang2024non,
  title={Non-Abelian physics in light and sound},
  author={Yang, Yi and Yang, Biao and Ma, Guancong and Li, Jensen and Zhang, Shuang and Chan, CT},
  journal={Science},
  volume={383},
  number={6685},
  pages={eadf9621},
  year={2024},
  publisher={American Association for the Advancement of Science}
}

@article{pachos2000optical,
  title={Optical holonomic quantum computer},
  author={Pachos, Jiannis and Chountasis, Spiros},
  journal={Physical Review A},
  volume={62},
  number={5},
  pages={052318},
  year={2000},
  publisher={APS}
}

@article{parto2023non,
  title={Non-Abelian effects in dissipative photonic topological lattices},
  author={Parto, Midya and Leefmans, Christian and Williams, James and Nori, Franco and Marandi, Alireza},
  journal={Nature Communications},
  volume={14},
  number={1},
  pages={1440},
  year={2023},
  publisher={Nature Publishing Group UK, London}
}

@article{chen2019non,
  title={Non-Abelian gauge field optics},
  author={Chen, Yuntian and Zhang, Ruo-Yang and Xiong, Zhongfei and Hang, Zhi Hong and Li, Jensen and Shen, Jian Qi and Chan, CT},
  journal={Nature communications},
  volume={10},
  number={1},
  pages={3125},
  year={2019},
  publisher={Nature Publishing Group UK, London}
}

@article{sun2022non,
  title={Non-Abelian Thouless pumping in photonic waveguides},
  author={Sun, Yi-Ke and Zhang, Xu-Lin and Yu, Feng and Tian, Zhen-Nan and Chen, Qi-Dai and Sun, Hong-Bo},
  journal={Nature Physics},
  volume={18},
  number={9},
  pages={1080--1085},
  year={2022},
  publisher={Nature Publishing Group UK, London}
}

@article{pinske2022symmetry,
  title={Symmetry-protected non-Abelian geometric phases in optical waveguides with nonorthogonal modes},
  author={Pinske, Julien and Scheel, Stefan},
  journal={Physical Review A},
  volume={105},
  number={1},
  pages={013507},
  year={2022},
  publisher={APS}
}

@article{neef2023three,
  title={Three-dimensional non-Abelian quantum holonomy},
  author={Neef, Vera and Pinske, Julien and Klauck, Friederike and Teuber, Lucas and Kremer, Mark and Ehrhardt, Max and Heinrich, Matthias and Scheel, Stefan and Szameit, Alexander},
  journal={Nature Physics},
  volume={19},
  number={1},
  pages={30--34},
  year={2023},
  publisher={Nature Publishing Group UK, London}
}

@article{lai2025non,
  title={Non-Abelian gauge field optics in the time domain},
  author={Lai, Yucheng and Zhang, Yongliang and Chang, Kai},
  journal={Physical Review Letters},
  volume={135},
  number={23},
  pages={233802},
  year={2025},
  publisher={APS}
}

@article{chen2025high,
  title={High-dimensional non-Abelian holonomy in integrated photonics},
  author={Chen, Youlve and Fan, Yunru and Larsonneur, Gulliver and Xiang, Jinlong and He, An and Wang, Guohuai and Zhang, Xu-Lin and Ma, Guancong and Zhou, Qiang and Guo, Guangcan and others},
  journal={Nature Communications},
  volume={16},
  number={1},
  pages={3650},
  year={2025},
  publisher={Nature Publishing Group UK, London}
}

@article{song2025artificial,
  title={Artificial gauge fields in photonics},
  author={Song, Wange and Yang, Yi and Lin, Zhiyuan and Liu, Xuanyu and Wu, Shengjie and Chen, Chen and Ke, Yongguan and Lee, Chaohong and Liu, Wei and Zhu, Shining and others},
  journal={Nature Reviews Physics},
  volume={7},
  number={11},
  pages={606--620},
  year={2025},
  publisher={Nature Publishing Group UK, London}
}

@article{pinske2022geometrically,
  title={Geometrically robust linear optics from non-Abelian geometric phases},
  author={Pinske, Julien and Scheel, Stefan},
  journal={Physical Review Research},
  volume={4},
  number={2},
  pages={023086},
  year={2022},
  publisher={APS}
}

@article{zhang2022non,
  title={Non-Abelian braiding on photonic chips},
  author={Zhang, Xu-Lin and Yu, Feng and Chen, Ze-Guo and Tian, Zhen-Nan and Chen, Qi-Dai and Sun, Hong-Bo and Ma, Guancong},
  journal={Nature Photonics},
  volume={16},
  number={5},
  pages={390--395},
  year={2022},
  publisher={Nature Publishing Group UK, London}
}

@article{cheng2025non,
  title={Non-Abelian lattice gauge fields in photonic synthetic frequency dimensions},
  author={Cheng, Dali and Wang, Kai and Roques-Carmes, Charles and Lustig, Eran and Long, Olivia Y and Wang, Heming and Fan, Shanhui},
  journal={Nature},
  volume={637},
  number={8044},
  pages={52--56},
  year={2025},
  publisher={Nature Publishing Group UK, London}
}

@article{brosco2021non,
  title={Non-Abelian Thouless pumping in a photonic lattice},
  author={Brosco, Valentina and Pilozzi, Laura and Fazio, Rosario and Conti, Claudio},
  journal={Physical Review A},
  volume={103},
  number={6},
  pages={063518},
  year={2021},
  publisher={APS}
}

@article{yang2019synthesis,
  title={Synthesis and observation of non-Abelian gauge fields in real space},
  author={Yang, Yi and Peng, Chao and Zhu, Di and Buljan, Hrvoje and Joannopoulos, John D and Zhen, Bo and Solja{\v{c}}i{\'c}, Marin},
  journal={Science},
  volume={365},
  number={6457},
  pages={1021--1025},
  year={2019},
  publisher={American Association for the Advancement of Science}
}

@article{yan2023non,
  title={Non-Abelian gauge field in optics},
  author={Yan, Qiuchen and Wang, Zhihao and Wang, Dongyi and Ma, Rui and Lu, Cuicui and Ma, Guancong and Hu, Xiaoyong and Gong, Qihuang},
  journal={Advances in Optics and Photonics},
  volume={15},
  number={4},
  pages={907--976},
  year={2023},
  publisher={Optica Publishing Group}
}

@article{polimeno2021experimental,
  title={Experimental investigation of a non-Abelian gauge field in 2D perovskite photonic platform},
  author={Polimeno, L and Fieramosca, A and Lerario, Giovanni and De Marco, Luisa and De Giorgi, Milena and Ballarini, Dario and Dominici, Lorenzo and Ardizzone, Vincenzo and Pugliese, Marco and Prontera, CT and others},
  journal={Optica},
  volume={8},
  number={11},
  pages={1442--1447},
  year={2021},
  publisher={Optical Society of America}
}

@article{neef2025non,
  title={Non-adiabatic holonomies for photonic quantum computing},
  author={Neef, Vera and Pinske, Julien and Wolterink, Tom AW and Becker, Karo and Heinrich, Matthias and Scheel, Stefan and Szameit, Alexander},
  journal={Optica Quantum},
  volume={3},
  number={1},
  pages={93--98},
  year={2025},
  publisher={Optica Publishing Group}
}

@article{kim2026programmable,
  title={Programmable lattices for non-Abelian topological photonics and braiding},
  author={Kim, Gyunghun and Li, Jensen and Piao, Xianji and Park, Namkyoo and Yu, Sunkyu},
  journal={Physical Review Letters},
  volume={136},
  number={4},
  pages={043804},
  year={2026},
  publisher={APS}
}

@article{iadecola2016non,
  title={Non-Abelian braiding of light},
  author={Iadecola, Thomas and Schuster, Thomas and Chamon, Claudio},
  journal={Physical review letters},
  volume={117},
  number={7},
  pages={073901},
  year={2016},
  publisher={APS}
}

@article{bialynicki2013role,
  title={The role of the Riemann--Silberstein vector in classical and quantum theories of electromagnetism},
  author={Bialynicki-Birula, Iwo and Bialynicka-Birula, Zofia},
  journal={Journal of Physics A: Mathematical and Theoretical},
  volume={46},
  number={5},
  pages={053001},
  year={2013},
  publisher={IOP Publishing}
}

@article{barnett2014optical,
  title={Optical dirac equation},
  author={Barnett, Stephen M},
  journal={New Journal of Physics},
  volume={16},
  number={9},
  pages={093008},
  year={2014},
  publisher={IOP Publishing}
}

@article{guo2021experimental,
  title={Experimental observation of non-Abelian topological charges and edge states},
  author={Guo, Qinghua and Jiang, Tianshu and Zhang, Ruo-Yang and Zhang, Lei and Zhang, Zhao-Qing and Yang, Biao and Zhang, Shuang and Chan, Che Ting},
  journal={Nature},
  volume={594},
  number={7862},
  pages={195--200},
  year={2021},
  publisher={Nature Publishing Group UK, London}
}

@article{sugawa2021wilson,
  title={Wilson loop and Wilczek-Zee phase from a non-Abelian gauge field},
  author={Sugawa, Seiji and Salces-Carcoba, Francisco and Yue, Yuchen and Putra, Andika and Spielman, IB},
  journal={npj Quantum Information},
  volume={7},
  number={1},
  pages={144},
  year={2021},
  publisher={Nature Publishing Group UK, London}
}

@article{silveirinha2017p,
  title={P{\textperiodcentered} T{\textperiodcentered} D symmetry-protected scattering anomaly in optics},
  author={Silveirinha, M{\'a}rio G},
  journal={Physical Review B},
  volume={95},
  number={3},
  pages={035153},
  year={2017},
  publisher={APS}
}

@article{silveirinha2015chern,
  title={Chern invariants for continuous media},
  author={Silveirinha, M{\'a}rio G},
  journal={Physical Review B},
  volume={92},
  number={12},
  pages={125153},
  year={2015},
  publisher={APS}
}

@article{geva2025polarization,
  title={Polarization inversion with parity--time-reversal--duality symmetric scatterers},
  author={Geva, Roee and Silveirinha, M{\'a}rio G and Kastner, Raphael},
  journal={Physical Review Letters},
  volume={135},
  number={5},
  pages={053802},
  year={2025},
  publisher={APS}
}

@article{peng2022phonons,
  title={Phonons as a platform for non-Abelian braiding and its manifestation in layered silicates},
  author={Peng, Bo and Bouhon, Adrien and Monserrat, Bartomeu and Slager, Robert-Jan},
  journal={Nature Communications},
  volume={13},
  number={1},
  pages={423},
  year={2022},
  publisher={Nature Publishing Group UK, London}
}

@article{fruchart2020dualities,
  title={Dualities and non-Abelian mechanics},
  author={Fruchart, Michel and Zhou, Yujie and Vitelli, Vincenzo},
  journal={Nature},
  volume={577},
  number={7792},
  pages={636--640},
  year={2020},
  publisher={Nature Publishing Group UK, London}
}

@article{yves2024symmetry,
  title={Symmetry-driven Phononic Metamaterials},
  author={Yves, Simon and Fruchart, Michel and Fleury, Romain and Shmuel, Gal and Vitelli, Vincenzo and Haberman, Michael R and Al{\`u}, Andrea},
  journal={arXiv preprint arXiv:2411.18556},
  year={2024}
}

@article{van2018photonic,
  title={Photonic Dirac monopoles and skyrmions: spin-1 quantization},
  author={Van Mechelen, Todd and Jacob, Zubin},
  journal={Optical Materials Express},
  volume={9},
  number={1},
  pages={95--111},
  year={2018},
  publisher={Optical Society of America}
}

@article{van2018quantum,
  title={Quantum gyroelectric effect: Photon spin-1 quantization in continuum topological bosonic phases},
  author={Van Mechelen, Todd and Jacob, Zubin},
  journal={Physical Review A},
  volume={98},
  number={2},
  pages={023842},
  year={2018},
  publisher={APS}
}

@article{van2021optical,
  title={Optical n-invariant of graphene’s topological viscous hall fluid},
  author={Van Mechelen, Todd and Sun, Wenbo and Jacob, Zubin},
  journal={Nature Communications},
  volume={12},
  number={1},
  pages={4729},
  year={2021},
  publisher={Nature Publishing Group UK, London}
}

@article{sun2023optical,
  title={Optical N-plasmon: topological hydrodynamic excitations in graphene from repulsive Hall viscosity},
  author={Sun, Wenbo and Van Mechelen, Todd and Bharadwaj, Sathwik and Boddeti, Ashwin K and Jacob, Zubin},
  journal={New Journal of Physics},
  volume={25},
  number={11},
  pages={113009},
  year={2023},
  publisher={IOP Publishing}
}

@article{kane2005quantum,
  title={Quantum spin Hall effect in graphene},
  author={Kane, Charles L and Mele, Eugene J},
  journal={Physical review letters},
  volume={95},
  number={22},
  pages={226801},
  year={2005},
  publisher={APS}
}

@article{qin2026photonic,
  title={Photonic exceptional points in engineered materials and their emerging applications},
  author={Qin, Haoye and Lv, Wenjing and Zhang, Zhe and Yang, Zijin and Li, Jue and Li, Mengyao and Li, Bo and Zhou, Ji and Fleury, Romain and Genevet, Patrice and others},
  journal={Nature Reviews Materials},
  volume={11},
  number={6},
  pages={453--468},
  year={2026},
  publisher={Nature Publishing Group UK, London}
}

\end{document}


\title{Supplemental Material for `Theorem for spin-1 photonic pairs and non-Abelian holonomy in Maxwell Hamiltonians'}

\author{Todd Van Mechelen}
\altaffiliation{equal contribution}
\affiliation{Elmore Family School of Electrical and Computer Engineering, Birck Nanotechnology Center, Purdue University, West Lafayette, Indiana 47907, USA}

\author{Wenbo Sun}
\altaffiliation{equal contribution}
\affiliation{Elmore Family School of Electrical and Computer Engineering, Birck Nanotechnology Center, Purdue University, West Lafayette, Indiana 47907, USA}

\author{Zubin Jacob}
\email{zjacob@purdue.edu}
\affiliation{Elmore Family School of Electrical and Computer Engineering, Birck Nanotechnology Center, Purdue University, West Lafayette, Indiana 47907, USA}

\maketitle 

\section{Normalized Riemann-Silberstein vectors} \label{sec:normalized_RS}
In this section, we present the Maxwell Hamiltonians following the normalized RS vectors. We demonstrate that normalized RS vectors extend the spin-1 photonic pair construction to material response with $\mu$ proportional to $\varepsilon$, and the Maxwell Hamiltonian of magnetoelectric media discussed in the main text remain unchanged under normalized RS vectors. In the following, we consider a homogeneous medium with $\mu$ proportional to $\varepsilon$, and the time reversal invariant magnetoelectric response with the anti-Hermitian part $i\gamma$ following constraints in Eq.~(10) in the main text. We define the normalized fields using the positive matrix square roots,
\begin{equation}
    \Psi_\pm
    =
    \frac{\sqrt{\varepsilon}\,\mathbf{E}
    \pm i\sqrt{\mu}\,\mathbf{H}}{\sqrt{2}},
    \qquad
    \Psi=
    \begin{pmatrix}
        \Psi_+\\
        \Psi_-
    \end{pmatrix}.
    \label{eq:sm_normalized_RS}
\end{equation}

Substituting Eq.~\eqref{eq:sm_normalized_RS} into Maxwell's equations gives
\begin{equation}
    \begin{aligned}
        \omega\mathcal{M}_{\mathrm{N}}\Psi
        &=
        \mathcal{H}_{\mathrm{N}}^0\Psi,\\
        \mathcal{M}_{\mathrm{N}}
        &=
        \mathcal{I}_6+\sigma_y\otimes
        \left(\varepsilon^{-1/2}\gamma\mu^{-1/2}\right),\\
        \mathcal{H}_{\mathrm{N}}^0
        &=
        c\,\sigma_z\otimes
        \left[
        \varepsilon^{-1/2}
        (\mathbf{k}\cdot\mathbf{S})
        \mu^{-1/2}
        \right].
    \end{aligned}
    \label{eq:sm_normalized_Maxwell}
\end{equation}
Proportionality between $\mu$ and $\varepsilon$ makes both $\varepsilon^{-1/2}\gamma\mu^{-1/2}$ and $\varepsilon^{-1/2}(\mathbf{k}\cdot\mathbf{S})\mu^{-1/2}$ Hermitian and purely imaginary. The normalized RS basis therefore retains $\mathcal{T}_{\mathrm{ph}}=K$ and $\widetilde{\mathcal{P}}_{\mathrm{ph}}=\sigma_y \otimes \mathcal{I}_3$, with
\begin{equation}
    \begin{aligned}
        \big[
        (\widetilde{\mathcal{P}}\mathcal{T})_{\mathrm{ph}},
        \mathcal{M}_{\mathrm{N}}
        \big]
        &=
        \big[
        (\widetilde{\mathcal{P}}\mathcal{T})_{\mathrm{ph}},
        \mathcal{H}_{\mathrm{N}}^0
        \big]
        =0,\\
        (\widetilde{\mathcal{P}}\mathcal{T})_{\mathrm{ph}}^2
        &=-\mathcal{I}_6.
    \end{aligned}
    \label{eq:sm_normalized_symmetry}
\end{equation}
Thus, the spin-1 photonic pairs are preserved. For scalar $\varepsilon,\mu>0$ and $\gamma=\gamma^z\hat{\mathbf{n}}\cdot\mathbf{S}$, Eq.~\eqref{eq:sm_normalized_Maxwell} reduces to the main text Maxwell Hamiltonian,
\begin{equation}\label{eq:sm_normalized_mass_Hamiltonian}
    \begin{aligned}
        \mathcal{H}_{\mathrm{ph}}^{\mathrm{ME}}\Psi &= v\left[
        \sigma_z\otimes(\mathbf{k}\cdot\mathbf{S})
        +
        m_{\mathrm{ph}}\sigma_y\otimes
        (\hat{\mathbf{n}}\cdot\mathbf{S})
        \right] \Psi
        =\omega\Psi,        
    \end{aligned}
\end{equation}
where $v=\frac{c}{\sqrt{\varepsilon\mu}}$, $m_{\mathrm{ph}}=-\frac{\omega}{c}\gamma^z$. This RS vector normalization thus leaves the main text effective mass definition unchanged.

For our model mass term $m_{\mathrm{ph}}=m_0-m_2k^2$, the energy metric in the normalized basis is
\begin{equation}
    \begin{aligned}
        \mathcal{W}_{\mathrm{N}}
        &=
        \partial_\omega
        \left(\omega\mathcal{M}_{\mathrm{N}}\right)
        =\mathcal{I}_6,\\
        \Psi^\dagger\Psi
        &=
        \varepsilon|\mathbf{E}|^2+\mu|\mathbf{H}|^2.
    \end{aligned}
    \label{eq:sm_energy_normalization}
\end{equation}
This justifies the energy inner product  we use to calculate its non-Abelian holonomy in the next section.

\section{Derivation of the non-Abelian Wilczek-Zee holonomy} \label{sec:WZ_derivation}
In this section, we present detailed derivations of the non-Abelian Wilczek-Zee holonomy of the massive spin-1 photonic pairs. Rotating the magnetoelectric axis changes the photonic eigenspace while preserving its twofold degeneracy. Following the main text, we consider $q=\sqrt{k^2+m_{\mathrm{ph}}^2}$, $a=k/q$, and $b=m_{\mathrm{ph}}/q$, with $k>0$ and $a^2+b^2=1$.

To construct the instantaneous pair $V=(\psi_{\mathbf{k}}^+,\psi_{\mathbf{k}}^-)$ during the adiabatic evolution, we introduce,
\begin{equation}
    \begin{aligned}
        \mathbf{u}(\varphi)
        &=
        \cos\varphi\,\hat{\mathbf{z}}
        -\sin\varphi\,\hat{\mathbf{t}}, \qquad
        \hat{\mathbf{n}}
        =
        \cos\theta\,\hat{\mathbf{k}}
        +\sin\theta\,\mathbf{u},\\
        \mathbf{e}_\theta
        &=
        \cos\theta\,\mathbf{u}
        -\sin\theta\,\hat{\mathbf{k}}, \qquad
        \mathbf{t}_\varphi
        =
        \sin\varphi\,\hat{\mathbf{z}}
        +\cos\varphi\,\hat{\mathbf{t}},
    \end{aligned}
    \label{eq:sm_orientation_frame}
\end{equation}
where $\hat{\mathbf{t}}=\hat{\mathbf{z}}\times\hat{\mathbf{k}}$ remains fixed. An orthonormal positive frequency pair is,
\begin{equation}
    \psi_{\mathbf{k}}^\zeta(\theta,\varphi)
    =
    \frac{1}{2}
    \begin{pmatrix}
        -a\mathbf{u}
        +\zeta b\mathbf{e}_\theta
        +i\mathbf{t}_\varphi\\
        -ib\mathbf{e}_\theta
        -i\zeta a\mathbf{u}
        +\zeta\mathbf{t}_\varphi
    \end{pmatrix},
    \qquad
    \zeta=\pm1.
    \label{eq:sm_instantaneous_pair}
\end{equation}
Direct substitution gives $\mathcal{H}_{\mathrm{ph}}^{\mathrm{ME}} \psi_{\mathbf{k}}^\zeta=vq\,\psi_{\mathbf{k}}^\zeta$. At $(\theta,\varphi)=(\pi/2,0)$,
Eq.~\eqref{eq:sm_instantaneous_pair} reproduces main text Eq.~(14).

For the instantaneous pairs $V=(\psi_{\mathbf{k}}^+,\psi_{\mathbf{k}}^-)$, in the adiabatic limit, we remove the common dynamical phase through $\Psi(t)\simeq e^{-ivqt}V(t)\mathbf{c}(t)$. Projection onto the doublet gives,
\begin{equation}
    \begin{aligned}
        \dot{\mathbf{c}}
        &=
        i\left(
        A_\theta\dot{\theta}
        +A_\varphi\dot{\varphi}
        \right)\mathbf{c}, \qquad
        A_\mu
        =iV^\dagger\partial_\mu V, \qquad
        W_C
        =
        \mathcal{P}_{\mathrm{path}}
        \exp\left(
        i\oint_C A_\mu\,d\lambda^\mu
        \right),
    \end{aligned}
    \label{eq:sm_WZ_evolution}
\end{equation}
where $\lambda^\mu=(\theta,\varphi)$. Using $\partial_\theta\mathbf{e}_\theta=-\hat{\mathbf{n}}$, $\partial_\varphi\mathbf{u}=-\mathbf{t}_\varphi$, $\partial_\varphi\mathbf{t}_\varphi=\mathbf{u}$, and $\partial_\varphi\mathbf{e}_\theta =-\cos\theta\,\mathbf{t}_\varphi$, we find from Eq.~\eqref{eq:sm_instantaneous_pair},
\begin{equation}
    A_\theta=\frac{ab}{2}\sin\theta\,\tau_y,
    \qquad
    A_\varphi=a\tau_x-b\cos\theta\,\tau_z.
    \label{eq:sm_WZ_connection}
\end{equation}
Here $\tau_i$ act on the ordered amplitudes $(c_+,c_-)^T$.

For the adiabatic evolution path $C_1$ in the main text, where $\theta=\pi/2$ and $\varphi$ follows $0\to2\pi$, we find the connection is,
\begin{equation}
    W_1=e^{i\Phi_1\tau_x},
    \qquad
    \Phi_1=2\pi a.
    \label{eq:sm_first_loop}
\end{equation}
Meanwhile for the adiabatic evolution path $C_2$ following $(\pi/2,0)\to(\theta_1,0)\to (\theta_1,2\pi)\to(\pi/2,2\pi)$, with $0<\theta_1<\pi/2$, we define $\delta=ab\cos\theta_{1}/2$. Integrating over the path, we find
\begin{equation}
    \begin{aligned}
        U_{\mathrm{out}}
        &=e^{-i\delta\tau_y}, \qquad
        U_{\mathrm{lat}}
        =e^{i2\pi(a\tau_x-b\cos\theta_{1}\tau_z)}, \qquad
        U_{\mathrm{ret}}
        =e^{i\delta\tau_y}, \qquad
        W_2
        =U_{\mathrm{ret}}U_{\mathrm{lat}}U_{\mathrm{out}}
        =e^{i\delta\tau_y}
        e^{i2\pi(a\tau_x-b\cos\theta_{1}\tau_z)}
        e^{-i\delta\tau_y}.
    \end{aligned}
    \label{eq:sm_second_loop}
\end{equation}
Expanding the matrix exponential gives,
\begin{equation}
    \begin{aligned}
        W_2
        &=
        \cos\Phi_2\,I_2
        +i\sin\Phi_2
        \left(u_x\tau_x+u_z\tau_z\right), \qquad
        \rho
        =\sqrt{a^2+b^2\cos^2\theta_{1}},
        \qquad
        \Phi_2=2\pi\rho,\\
        u_x
        &=
        \frac{
        a\cos(2\delta)+b\cos\theta_{1}\sin(2\delta)
        }{\rho}, \qquad
        u_z
        =
        \frac{
        a\sin(2\delta)-b\cos\theta_{1}\cos(2\delta)
        }{\rho}.
    \end{aligned}
    \label{eq:sm_second_loop_axis}
\end{equation}
The two closed cycles therefore satisfy
\begin{equation}\label{eq:sm_loop_commutator}
    \begin{aligned}
        [W_1,W_2]
        &=
        2i\sin\Phi_1\sin\Phi_2\,u_z\tau_y\neq 0.
    \end{aligned}
\end{equation}
A change of basis at their common starting point conjugates both $W_1$ and $W_2$ by the same unitary matrix. Their noncommutativity is therefore basis independent. Equation~\eqref{eq:sm_loop_commutator} establishes non-Abelian holonomy whenever its right hand side is nonzero. Meanwhile, we note some special loop parameters can make it vanish.

For the initial state $\psi_{\mathbf{k}}^+$ discussed in the main text, let $V_0=V(\pi/2,0)$ and $\mathbf{c}_0=(1,0)^T$. The two adiabatic evolution orders give,
\begin{equation}
    |f\rangle=V_0W_2W_1\mathbf{c}_0,
    \qquad
    |f'\rangle=V_0W_1W_2\mathbf{c}_0.
    \label{eq:sm_ordered_outputs}
\end{equation}
At $\theta_1=\pi/3$, from Eq.~\eqref{eq:sm_ordered_outputs}, we find $|\langle f|f'\rangle|^2 \approx 0.85$, matching the distinct final states in the main text Fig.~3.


\bibliography{reference.bib}